\documentstyle[psfig]{mn}

\input{epsf}

\begin{document}
\title[Shapes of clusters and groups of galaxies]
  {Shapes of clusters and groups of galaxies: Comparison of model predictions with observations}

\author[Paz, Lambas, Padilla \& Merch\'an]{
\parbox[t]{\textwidth}{
D. J. Paz$^{1,2\ \star}$,
D. G. Lambas$^{1,2}$, 
N. Padilla$^3$, 
M. Merch\'an$^{1,2}$}
\vspace*{6pt} \\ 
$^1$ Grupo de Investigaciones en Astronom\'{\i}a Te\'{o}rica 
    y Experimental (IATE), Observatorio Astron\'omico de 
    C\'{o}rdoba, UNC, Argentina.\\
$^2$ Consejo Nacional de Investigaciones Cient\'{\i}ficas y Tecnol\'ogicas
    (CONICET), Argentina.\\
$^3$ Departamento de Astronom\'\i a y Astrof\'\i sica, Pontificia
     Universidad Cat\'olica, V. Mackenna 4860, Casilla 306, Santiago 22, Chile.\\
$^{\star}$ e-mail: dpaz@oac.uncor.edu\\
}

\maketitle
\begin{abstract}
We study the properties of the 3-dimensional and projected shapes of haloes
using high resolution numerical simulations and observational data where the 
latter comes from the 2PIGG (Eke et al. 2004) and SDSS-DR3GC group catalogues 
(Merch\'an \& Zandivarez 2005).  We investigate the dependence of halo shape on 
characteristics such as mass and number of members. In the 3-dimensional case, 
we find a significant correlation between the mass and halo shape; massive 
systems are more prolate than small haloes.
We detect a source of strong systematics in estimates of the triaxiality
of a halo, which is found to be a strong function of the number
of members; $\Lambda$CDM haloes usually characterised by triaxial shapes, slightly bent 
toward prolate forms,  appear more oblate when
taking only a small subset of the halo particles. \\
The ellipticities of observed 2PIGG and SDSS-DR3GC groups are found to be strongly 
dependent on the number of group members, so that poor groups appear more elongated
than rich ones.  However, this is again an artifact caused by poor statistics
and not an intrinsic property of the galaxy groups, nor an effect from
observational biases.
We interpret these results with the aid of a GALFORM mock 
2PIGG catalogue.  When comparing the group
ellipticities in mock and real catalogues, we find an excellent agreement
between the trends of shapes with number of group members.
When carefully taking into account the effects of low number statistics,
we find that more massive groups are consistent with more elongated shapes.
Finally, our studies find no significant correlations between the shape
of observed 2PIGG or SDSS-DR3GC groups with the properties of galaxy members
such as colour or spectral type index.
\end{abstract}


\section{Introduction}

Cluster and groups of galaxies provide invaluable information on the
formation and evolution of structure in the Universe. These systems, 
also denominated haloes, represent the largest gravitationally bound systems 
in the universe.  It has been shown that these systems are mildly
aspherical with orientations related to the surrounding structures such
as filaments and large scale walls (see Kasun \& Evrard, 2005,
and references therein).  Results from numerical simulations
by van Haarlem \& van de Weygaert (1993, see also Splinter et al., 1997) 
have shown that the origin for such alignments comes from re-arrangements
of the halo axes in the direction of the accretion anisotropy (e.g.
last major merger event).  Therefore, in a statistical sense, the
shapes of haloes could encode information about the large-scale
structure in the Universe.  On the other hand, current halo 
models (Cooray \& Sheth, 2002, van den Bosch et al. 2004), which
successfully describe several galaxy statistics, assume that 
the distribution of mass in haloes is spherical.  However, a more
accurate version of the model would need to take into account the
actual complicated internal structure (Jing \& Suto, 2002),
and aspherical shapes of haloes.

Halo properties have been studied extensively using numerical
simulations. For instance, high resolution, but relatively
small volume simulations have been used to provide detailed information on
the halo density profile and halo shape, whereas larger
volume, but low resolution simulations (e.g. the Hubble Volume simulations,
Evrard et al. 2002) have provided information on the abundance and spatial
distribution of haloes (Colberg et al., 1999, ,Colberg et al. 2000,
Jenkins et al., 2001, Padilla \& Baugh, 2002).

Several authors (Warren et al., 1992, Thomas et al., 1998, Hopkins, Bahcall
\& Bode, 2005, Kasun \& Evrard, 2005, and references there in)
have analysed the distribution of halo shapes using the best fitting 
ellipsoid to the spatial distribution of halo members.  They calculate
the minor to major
semi-axis ratio of each halo, and use it as an indicator of the best-fitting ellipsoid shape.
In order to obtain reliable distributions of semi-axis ratios, it
is necessary to reach an equilibrium between spatial resolution
(to improve the estimate of semi-axis ratios) and comoving volume 
(for good statistics).
Kasun \& Evrard (2005) combine large volume and high resolution simulations,
to find a systematic trend with halo shape: a larger ratio of minor to major semi-axes
is obtained when larger halo masses are considered.
These authors also find good alignments between velocity 
and spatial principal semi-axes.
Furthermore they find that cluster shapes are independent of 
environment and reflect the filamentary structure of the universe through
a non-random alignment at very large scales of up to 200 $h^{-1}$ Mpc.

On the observational side,
Plionis, Basilakos, \& Tovmassian (2004) find a trend of shape with
cluster size that is opposite to that seen in simulations 
(Kasun \& Evrard, 2005).
From 1168 groups in the UZC-SSRS2 galaxy group catalogue, Plionis et al.
concluded that poor groups are more elongated than rich ones, with $85\%$ 
of poor groups having a projected semi-axis ratio lower than $0.4$.\\

In this work, we explore shapes of dark matter haloes with masses ranging 
from groups to clusters of galaxies using a high resolution cosmological simulation
with a modest volume ($250^3 $h$^{-3}$Mpc$^3$).  We analyse the 
dependence of triaxiality, asphericity and projected shape 
on group mass, and number of members.
By studying observational samples of groups from the 2PIGG (Eke et al. 2004,
constructed from the 2-degree Field Galaxy Redshift Survey, Colless et al. 2001) 
and the SDSS-DR3GC group catalogue (Merch\'an \& Zandivarez, 2004, constructed
using the Data Release 3 of the Sloan Digital Sky Survey, SDSS-DR3,
Abazajian et al., 2004), we will demonstrate that the apparent inconsistency between the 
observational and simulated groups is mainly due to
an artifact produced by low number statistics. To study this apparent 
discrepancy we combine three-dimensional and two-dimensional analyses 
performed over a numerical simulation box and a mock catalogue, respectively.
In both cases the cosmology corresponds to a standard $\Lambda$CDM model.  In addition,
when constructing the mock catalogue, we make use of the
GALFORM (Cole et al. 2000) semi-analytic galaxies populating the simulation box.
In particular, the mock 
catalogue provides us with an excellent test sample to explore the 
outcomes of group and cluster shape measurements
within the current cosmological context.

This paper is organised as follows.  In section 2 we provide a study of
3-dimensional shapes of haloes in the numerical simulation, including
analyses of systematic biases coming from low number statistics.  Section 3 contains
a study of projected group shapes from observational datasets,
and provides a theoretical framework to interpret our results by means of
a mock group catalogue and the projection onto two dimensions of the 3-dimensional
haloes in the simulation.  Finally, we summarise the main conclusions from this
work in section 4.

\section{3-Dimensional Analysis}

In this section we analyse the predicted shape of gravitationally bound  
systems with masses ranging from groups to clusters of galaxies 
($~10^{12}$h$^{-1}M_\odot$ to $~10^{15}$h$^{-1}M_\odot$). These systems were taken
from a cosmological numerical simulation kindly provided by
the Durham group, performed using the first version
of the GADGET code developed by Springel et al. (2001). The computational volume 
corresponds to a periodic box of side $250$h$^{-1}$Mpc containing $500^3$ 
particles with masses $M=1.04\times10^{10} $h$^{-1}M_\odot$. The simulation adopts 
a Cold Dark Matter
(CDM) model with density parameters
 $\Omega_0=0.3$, $\Omega_\Lambda=0.7$, and a present $rms$
mass fluctuations of $\sigma_8=0.8$. 
The identification of clumps of particles is carried out by means of a
standard friends-of-friends algorithm, with a percolation length of 
$l=0.17n^{-1/3}$, were $n$ is the mean number density.  We only consider haloes
with at least $10$ particles.

\subsection{Shapes of Halos}

\begin{figure}
{\epsfxsize=8.truecm 
\epsfbox[40 170 575 685]{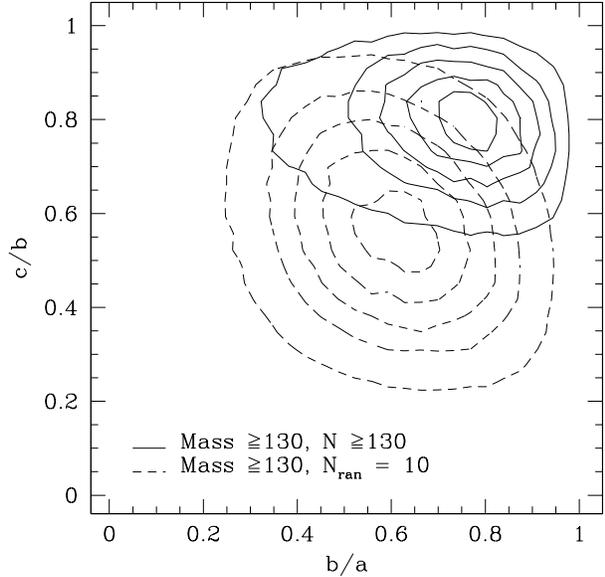}}
\caption{
   Contour maps of the scatter-plot of $b/a$ vs. $c/b$ semi-axis ratios (see text) 
   for groups with
   masses $M>10^{12}M_\odot$. Solid lines correspond to the values estimated
   using all members, while dashed lines show $b/a$ vs. $c/b$ when 
   only 10 random group members are used in the calculation.  Contours
   enclose $10$, $30$, $50$, $70$, and $90\%$ of the $b/a$ vs. $c/b$
   points (inner to outer contours, respectively).}
\label{fig:scatprol}
\end{figure}

For each dark-matter halo, we calculate the inertia tensor using positions of halo
members.  This can be written as a symmetric matrix,
\begin{equation}
I_{ij}= (1/N_h)\sum_{\alpha=1}^{N_h} X_{\alpha i} X_{\alpha j},
\end{equation}
where $X_{\alpha i}$ is the $i^{th}$ component of the displacement vector
of a particle $\alpha$ relative to centre of mass, and $N_{h}$ is the number
of particles in the halo. The matrix eigenvalues correspond to the square of the semi-axis 
($a$, $b$, $c$ were $a>b>c$) of the characteristic ellipsoid
that best describes the spatial distribution of the halo 
members. Our group shape analysis is based on the semi-axis
ratios $b/a$ and $c/b$. These variables are independent and provide a 
complete set of parameters to analyse the ellipsoid shape. Several authors 
use the quotient of minor to major eigenvalues ($c/a$), since
this ratio provides more appreciable changes with the ellipsoid asphericity.
However, such a
parameter does not discern between oblate and prolate ellipsoids.
Figure 1\  
shows the contour map of the scatter plot of $b/a$ vs $c/b$
ratios for groups with masses $M>10^{12}M_\odot$ (solid lines containing
$10$, $30$, $50$, $70$, and $90\%$ of the $b/a$ vs. $c/a$ pairs).
Here a fixed $b/a=1$ with an arbitrary value of $c/b$, corresponds to 
perfect oblate ellipsoids. On the other hand, systems with fixed $c/b=1$ are
perfect prolate ellipsoids. A system with $b/a<c/b$
is associated to a general triaxial ellipsoid with prolate tendency, while the opposite
case, $b/a>c/b$, corresponds to a predominantly oblate ellipsoid.
As can be seen, haloes are generally triaxial; their shapes are almost uniformly
distributed between the two extremes (inner contours) with a slight
preference (outer contours) to prolate configurations (i.e. $a\gg b>c$).
The most frequent values of semi-axis ratios are $c/b=0.8$ and $b/a=0.76$.
This is in qualitative agreement with Frenk et al. (1998) but with much higher
statistical significance.

\subsection{Resolution Effects in the determination of halo shapes}

When using galaxy catalogues and low resolution simulations, only a few group members
are available to estimate the inertia tensor.  We therefore study possible low number
effects in the estimate of semi-axis ratios. 
There are two main effects expected to take part in biasing the distribution of
shapes.  On the one hand, semi-axes of similar amplitude will be affected by shot noise
due to the use of a small number of discrete particles when calculating the inertia tensor.
This will induce a rearrangement of the semi-axes resulting in smaller $c/b$ and
$b/a$ semi-axis ratios.  On the other hand, the inertia tensor is a second order
measure, which implies that a further, more complicated bias is to be expected. 
We test these hypotheses by
computing halo semi-axis ratios using only a subsample
of randomly selected halo members. Figure 1 
(dashed lines) shows 
iso-density contours of $b/a$ vs. $c/b$ ratios determined using 10 randomly 
selected particles.  When comparing with the results
obtained using all the halo particles it can be noticed that the shape parameters
from all members have been shift towards lower $c/b$ and $b/a$ values, in
agreement with our first hypothesis.  In addition, 
it can be seen that the degradated distribution is shifted toward 
more oblate values with a peak density at lower $b/a$ and $c/b$ 
ratios. 
We identify this systematic effect with our second source of biases.
The reason for this phenomenon could reside in the fact that 
shot noise can only reduce the signal (ie. by definition: $b/a <1$ and $c/b<1$), which
makes the smaller semi-axis more prone to be affected by noise. 
Therefore, prolate shapes (two small semi-axis) are more distorted by noise than
oblate shapes (only one small semi-axis).\\
To probe quantitatively these concepts, we perform a Monte-Carlo simulation
and homogeneously cover
the parameter space (b/a, c/b) with ellipsoids populated with $10,000$ points,
and calculate average semi-axis ratio estimates, determined
using sets of different numbers of randomly selected ellipsoid points.
Figure 2 
shows the path (solid lines) described by the ellipsoid shapes as
the number of members used to calculate the inertia tensor 
(hereafter $N_{ran}$) is reduced from $1000$ to $5$. The open squares show the underlying values 
for each Monte-Carlo ellipsoid.
The filled circles show the results when only five randomly extracted points
are used to calculate $b/a$ and $c/b$.
As we have mentioned previously, the effect of the low number statistics 
is to produce a systematic displacement to lower values of $b/a$ and $c/b$, 
with a stronger effect on 
spherical shapes with a tendency to populate the oblate region.

\begin{figure}
{\epsfxsize=8.truecm 
\epsfbox[40 170 575 705]{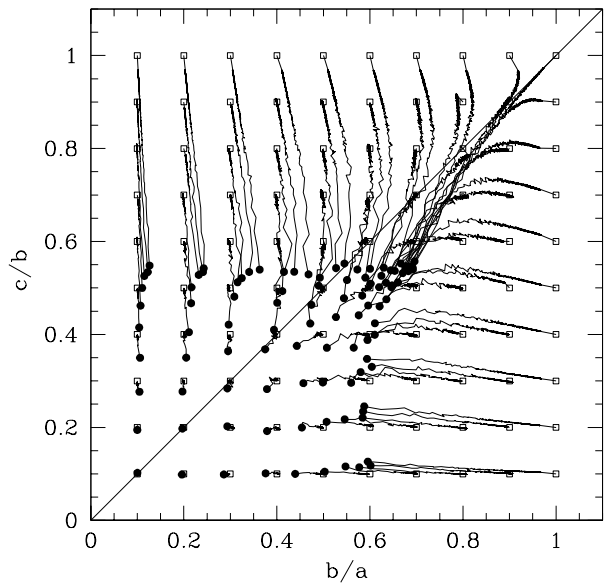}}
\caption{
  Dependence of semi-axis ratio estimates ($b/a$ and $c/b$) on the
           number of randomly selected points used to calculate the inertia 
           tensor (solid lines). The open squares are the set values in each 
           Monte-Carlo ellipsoid; the filled circles represent the values obtained
           using sets of five randomly extracted points from each Monte-Carlo ellipsoid.}
  \label{fig:montecarlo}
\end{figure}

\subsection{Shape dependence on cluster mass}

\begin{figure}
{\epsfxsize=8.truecm 
\epsfbox[40 170 575 705]{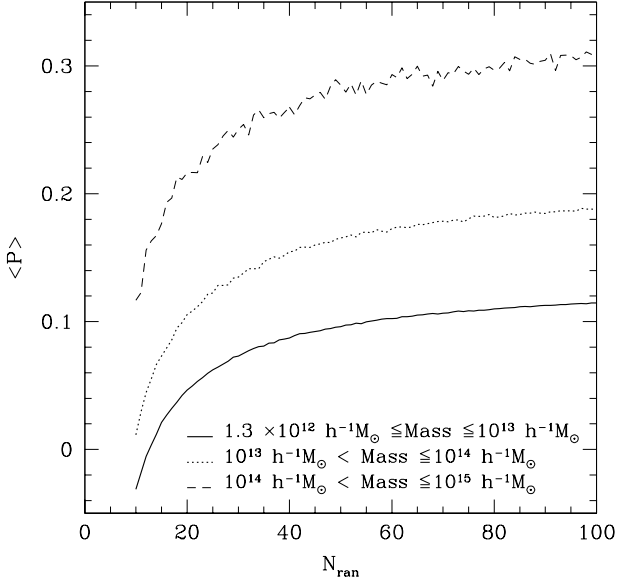}}
  \caption{The evolution of the triaxiality parameter $P$ with $N_{ran}$ for
           ranges of increasing halo mass (solid, dotted, and dashed lines
	   respectively).}
  \label{fig:prolcurves}
\end{figure}

The analysis in the previous section showed that haloes are preferentially
prolate.  Previous works 
(Kasun \& Evrard, 2005, Jing \& Suto, 2002) are consistent with a weak dependence of the 
mean $c/a$ parameter on mass, which is commonly used to indicate 
the asphericity of a dark-matter halo.  In order to provide a suitable 
characterization of the shape dependence on mass and number of particles,
we introduce a new parameter that allows us to differentiate between prolate
and oblate ellipsoids, which we find more suitable to perform the task
than  the $c/a$ ratio.  A simple way to define this parameter,
is to calculate the quotient between the parameters $c/b$ and $b/a$; 
prolate systems satisfy $ca/b^2>1$, whereas oblate systems, $ca/b^2<1$.
Consequently we define the triaxiality parameter as:
\begin{equation}
 P:= \ln(ca/b^2).
\end{equation}
Using the logarithm avoids the natural skewness of the $ca/b^2$ distribution,
arising from the asymmetry between the oblate ($0<ca/b^2<1$) and prolate
($1<ca/b^2< \infty$) parameter ranges. Therefore an oblate (prolate) ellipsoid
satisfies $P<0$ ($P>0$).
Regardless of the number of particles used to calculate the inertia tensor, 
we find a dependence of the distribution of cluster shapes on mass
consistent with previous determinations.
Figure 3
shows the triaxiality parameter as a function of
number of particles $N_{ran}$ for different mass ranges.
In spite of the tendency of increasing values of $P$ with $N_{ran}$
(which reflects the number effect described in the previous section)
curves corresponding to different mass ranges are well separated 
indicating the triaxiality dependence on cluster mass regardless of $N_{ran}$.

\begin{figure*}
{\epsfxsize=16.truecm 
\epsfbox[18 430 592 718]{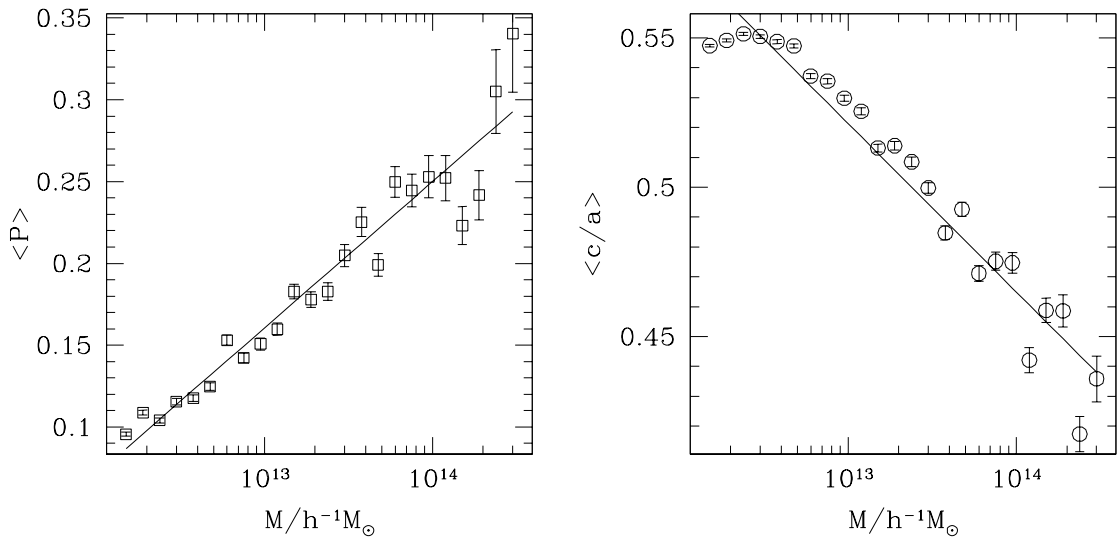}}
\label{fig:massco}
\caption{Triaxiality parameter $P$, and $c/a$ ratios as a function of halo mass (symbols in left
and right panels respectively).  The lines show the best fitting log-linear relations 
to the measurements from the numerical simulation (Equations 3 and 4.)}
\end{figure*}

In Figure 4\ 
we show the dependence of the mean values of the triaxiality
parameter $P$ (left panel) and $c/a$ ratio (right panel) on mass. Error bars correspond to
the standard $rms$ error in $P$ and $c/a$. Consistent with the results for $c/a$ obtained by
Kasun \& Evrard (2005) and Jing \& Suto (2002), 
we find a decreasing log-linear relation between the $c/a$
ratio and halo mass (i.e. the asphericity increases). We also find that $P$ increases 
with mass, indicating that the increment in asphericity comes from an increased prolaticity in 
the halo population.  The fitting formulae shown by the solid lines correspond to,
\begin{equation}
<c/a>=-(0.056 \pm 0.003)\log_{10}(M)+(0.70 \pm 0.01),
\end{equation}
\begin{equation}
<P>=(0.089 \pm 0.005)\log_{10}(M)-(0.11 \pm 0.01).
\end{equation}
The trend in $c/a$ is significantly more ellipsoidal than the results from Kasun \& Evrard (2005),
but as was pointed out by these authors in relation to the work by Thomas et al. (1998),
the reason for such discrepancies lies in the definition of halo shape. 
The boundary constraint of the spherical over-density method used 
by these authors to identify haloes, 
tends toward rounder measures of the inertia moments, whereas our haloes
tend to be more elongated due to the directional nature of the percolation process and the
pruning in local density. As will be shown later in this work, the trend in projected
semi-axis ratios with mass is that of more elongated shapes for more massive haloes.

The results obtained in this section indicate that even though there is
a clear tendency in triaxiality and asphericity with halo mass, statistical
biases coming from the number of halo members used to calculate
the halo shape, must be carefully considered.  Our analyses suggest that 
in order to obtain a reliable
dependence of halo shape properties with mass one should consider using a fixed
number of halo members regardless of halo mass.  This can certainly result in
a systematic off-set in the shape-mass relation, but will allow the detection
of any underlying trend with mass.

\section{Projected Analysis}

Due to redshift distortions, the analysis of observational data from redshift surveys
is limited to the measurement of group shapes as seen projected
on the plane of the sky.  Therefore, when studying observational samples of groups,
we estimate projected shapes following a similar procedure 
to the one described for the 3-dimensional case.
Using the projected Cartesian coordinates on the plane of the sky, we calculate
the 2-dimensional inertia tensor, whose eigenvalues provide the semi-axes $a$ and $b$
(major and minor semi-axes, respectively).
In this work we analyse the projected shapes of groups in the 2PIGG (2dFGRS Percolation
Inferred Galaxy Groups), constructed from the full 2-degree Field
Galaxy Redshift Survey (2dFGRS) by Eke et al. (2004), and in the
group catalogue compiled by Merch\'an \& Zandivarez (2005) from the Data Release 3 of 
the Sloan Digital Survey (SDSS-DR3GC).  The 2PIGG catalogue contains $4,045$ 
groups with at least $4$ members, and the SDSS-DR3GC group catalogues, $10,152$ groups.
In order to compare observational and numerical simulation results we construct 
a mock 2dFGRS catalogue by placing an observer at random within 
the simulation box, and reproducing as many of the 2dFGRS observational biases
as possible.
This is done by placing the same angular completeness mask 
and selection function as in the real 2dFGRS survey, and by measuring the 
distances to galaxies using redshifts instead of coordinate distances
(For more details on 2dFGRS mock construction see Eke et al., 2004).
The galaxies in the mock catalogues correspond to GALFORM semi-analytic 
galaxies populating the numerical simulation already described in earlier sections.    
After generating the mock galaxy catalogue, 
we apply the group finding algorithm originally used to identify the group 
catalogues in the real data, namely the Eke et al. (2004) Friends-of-Friends 
(FOF) algorithm to the 2dFGRS mock catalogue (software kindly provided by Vincent Eke).



\subsection{Dependence of group shape on number of members}

The study of halo shapes in the numerical simulation shows that
results can change dramatically when different numbers of halo
members are used to infer the halo shape.  In this section
we will assess the importance of this effect when calculating
the 2-dimensional inertia tensor of groups identified from
a galaxy redshift survey.

In order to do this, we proceed to calculate the shapes of groups in the real
and mock 2PIGG catalogues and compute the
distribution of shapes for different numbers of group members, $N_{gal}$.
We show the results from this calculation in
Figure 5.\ \ 
As can be seen, there is a striking resemblance 
between the results from mock and real data.  For instance, in both cases
the distribution obtained from groups with $N_{gal}\ge 4$ corresponds to more elongated 
projected shapes than that corresponding
to $N_{gal}\ge 20$. The agreement for different values of $N_{gal}$ between
data and mock groups is excellent, specially for the low $N_{gal}$
samples.  This indicates that the Cole et al. (2000)
model is adequate for this comparison.  It is also interesting
to observe the apparent increase of the typical semi-axis ratio with the number of group members.
As we will later demonstrate, this observed trend is mainly due to low number statistics.
The same figure also shows the results for groups in the SDSS-DR3GC, which are
in excellent agreement with mock and real 2PIGG group shapes.
This is a good indication of the robustness of our results, since
the group identification algorithms used in the 2PIGG and SDSS-DR3GC group
catalogues are slightly different,
the catalogues cover different areas on the sky, have slightly different depths, 
and suffer from very different observational biases due to the instrument
setups used in the construction of the surveys (see Colless et al., 2001, 
for the 2dFGRS and Abazajian et al., 2004, for the SDSS-DR3).

\begin{figure}
{\epsfxsize=8.truecm 
\epsfbox[40 170 575 705]{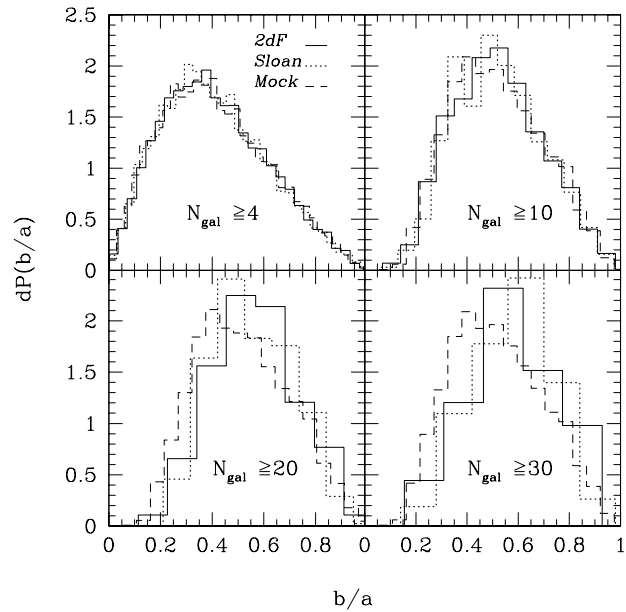}}
  \label{fig:ba.2df2sloan}
  \caption{Distributions of $b/a$ for the 2PIGG (solid lines), mock 2PIGG (dashed
lines) and SDSS-DR3GC (dot-dashed lines) groups, for different ranges of $N_{gal}$ 
(different panels, see figure keys).}
\end{figure}

The increase in the average values of $b/a$ with the number of group members
seen in Figure 5, is in agreement with the study of Plionis et al. (2004)
of the UZC-SSRS2 groups (see above).  There have been suggestions
for explaining the variation in typical group shapes with
$N_{gal}$, such as mechanisms for the accretion of matter which
would depend on the halo mass, or dynamical friction (Plionis et al., 2004).
However, studies of haloes in numerical simulations seem to point
in the opposite direction (Section 2, this work,
Kasun \& Evrard, 2004, Hopkins et al., 2005),
which poses a controversy between theory and observations.  A solution
to this discrepancy was proposed by Kasun \& Evrard (2004), whereby
the optical selection of galaxies would be responsible for important biases
that would reverse the dependence of halo shape on mass.  

We now search for the reason behind the increase in the typical 
projected semi-axis ratio with the number of group members, $N_{gal}$.
Our results from the study of halo shapes in the
numerical simulation have shown that the use of a low number of halo members
tends to lower the halo triaxiality parameter in a systematic way
(see Section 2).  We apply
the same procedure to the observational data, and compare the
distribution of semi-axis ratios for groups with $N_{gal}=6$ to the
distribution measured from groups with $N_{gal}\ge 20$ members but only
considering $6$ randomly selected galaxies from each group
(for example, out of a group with $20$ members, one can select
$38760$ subgroups of $6$ members each; we present our results from 
calculating the shapes of only $200$ subgroups per individual group with
$N_{gal}>19$, this number of subgroups ensures statistical independence to
some degree).  Figure 6\ 
shows the distribution of group semi-axis ratios with low values
of $N_{gal}=6$ (long-dashed histogram) and compares it to the distribution 
of groups with $N_{gal}>19$ resulting from using all the group members
(solid histogram; again, groups with larger number of members are consistent
with rounder projected shapes) and when selecting only $6$ members for 
measuring their shapes (dot-dashed histogram).
It is clear from this figure that the distribution of
group shapes with $N_{gal}=6$ is indistinguishable from that
of richer groups where only $6$ members are considered when
measuring their shapes (apparently more elongated than when using all
the group members).  The latter distribution is very smooth due to the large
number of subgroups considered.  
Therefore, we can conclude that the trend of shape
with the number of group members is mainly due to a systematic effect, and
that it may very well be possible that intrinsic group shapes do not vary significantly
with the number of group members.  

\begin{figure} {\epsfxsize=8.truecm 
\epsfbox[40 170 575 705]{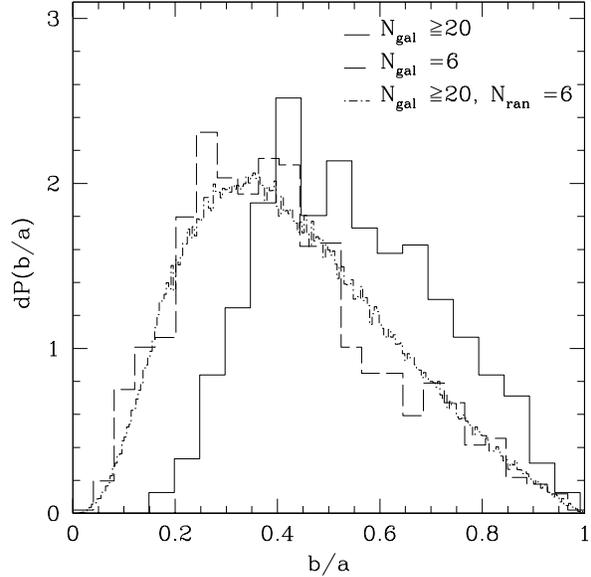}}
\label{fig:comp6y20}
\caption{Distribution of projected semi-axis ratios $b/a$ from groups
with $N_{gal}=6$ (dashed lines), groups with $N_{gal}>19$ (solid lines)
and groups with $N_{gal}>19$ where only $6$ group members were used
to obtain the group shape (dot-dashed lines).}
\end{figure}

Based on these results, we provide a measurement of an
unbiased distribution of group shapes from groups with any number
of members (bearing in mind that any dependence of the group shape on
mass will be averaged out in this analysis).   
In order to do this, we generate samples of subgroups
with numbers of members ranging from $4$ to $16$, extracted
from a sample of groups with $N_{gal}>19$ ($200$ subgroups from
each group), and calculate the distribution of shapes for the
subgroups of different numbers of members.  We fit a $4$th order
polynomial to the result using $N_{ran}$ extracted members,
\begin{equation}
F_{N_{ran}}(b/a)=\sum_{i=0}^4 h_i (b/a)^i.
\end{equation}
We then define correction factors which
can be applied to the distributions of semi-axis ratios of groups
with $N_{gal}$ members, and make these measurements equivalent
to what would result from using $\ge 20$ members instead of $N_{gal}$,
\begin{equation}
C_{N_{gal}}(b/a)=F_{\ge20}(b/a)/F_{N_{ran}=N_{gal}}(b/a),
\end{equation}
where $F_{\ge20}$ is the fit to the measured distribution of
groups with $N_{gal}\ge20$ members.
Note that the correction factors are obtained from the same
sample of groups to which these will be applied.
We show in Figure 7\ 
the resulting corrected
distributions of semi-axis ratios for $N_{gal}$ ranging
from $4$ to $13$.  As can be seen, the corrected distributions
for different $N_{gal}$ show some uncorrelated scatter around
a single distribution that can be fitted by a
fourth order polynomial.  For the 2PIGG catalogue,
the best fitting coefficients are $h_0=0.22\pm0.07$,
$h_1=-7\pm1$, $h_2=37.9\pm4.5$, $h_3=-49\pm7$, and $h_4=19.5\pm 3$.
As can be seen in this figure, the best fit polynomials corresponding
to the results from the 2PIGG, mock 2PIGG and SDSS-DR3GC groups are
extremely similar.

\begin{figure}
{\epsfxsize=8.truecm 
\epsfbox[40 170 575 705]{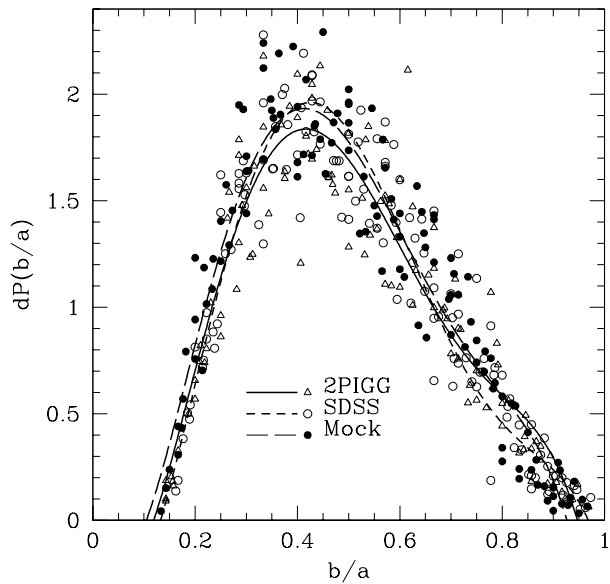}}
  \caption{Distributions of projected $b/a$ semi-axis ratios corrected to an
           equivalent of $N=20$ group members used for computing the group shapes.
           Results are shown for the 2PIGG, mock 2PIGG and SDSS-DR3GC catalogues
           (see the figure key).}
  \label{fig:corr}
\end{figure}

\subsection{Dependence of group shapes on mass}

In order to avoid systematic trends in the shapes of haloes, we
devise a simple statistical method that will unveil any shape
dependence on mass in the observational data.

As a first step to provide a connection between the results from the 
3-dimensional shapes of haloes in the full simulation box we simply project 
onto two dimensions a sample of haloes taken from the simulation.  This
sample is characterised by the same mass 
distribution present in the 2PIGG catalogue, ensuring the study of similar samples of
groups from the simulation and the data.  The masses of 2PIGG groups are
calculated using group luminosities and a $M/L$ relation from
Eke et al. (2004).
To achieve a realistic shape comparison, the number effect is also
taken into account by forcing the same distribution of number of members
present in the 2PIGG catalogue (we do this by randomly selecting members 
from each group in the simulation).  The distributions of projected  semi-axis 
ratios $b/a$ of this particular halo sample (solid line) and the mock 
2PIGG catalogue (doted line) are shown in Figure 8.\ \ 
As can be seen, both distributions are in reasonably good agreement.
However, mock groups show a weak tendency
to be more elongated than those obtained by direct projection
of the simulations.  Even though this effect is small,
we find that it persists even when sampling different
ranges of mass and number of members.  Therefore, 
it is possible that the offset shown by both distributions is due
to differences between the identification
algorithms used on simulations and redshift-space samples. For instance,
the latter algorithms use a complicated linking volume, characterised
by an assumed angular to radial aspect ratio, and a scaling with redshift
(see Eke et al., 2004, for more details).  This comparison could be adopted
in future works as a further test for assessing the quality of identification 
algorithms applied to redshift data.

As previously mentioned, a good way to avoid the number effect on
our statistics is to use a fixed number of group members to calculate 
the group shapes.  We therefore compute the average $b/a$ ratios of
projected dark-matter haloes as a function of halo mass, 
using only $10$ halo members out of the total available particles in each halo.  
The resulting semi-axis ratios can be seen in Figure 9,\ 
where more massive haloes are consistent with slightly more elongated shapes
than less massive haloes (open circles).  By comparing this trend with
halo mass with what is obtained using
all the halo particles for haloes with more than $100$ members, we find only
a constant systematic shift between the two.  This indicates that the procedure
of selecting only $10$ halo members preserves the trend of projected shape
with mass.  We now proceed to calculate the average $b/a$ ratios of groups
in the mock 2PIGG catalogue, for all groups with at least $10$ galaxy members,
but using only $10$ randomly selected members to calculate the group shape.  We
plot this result as a function of group mass, calculated using the $M/L$ ratio
characterising the mock catalogue (open squares).  As can be seen, the trend
with mass is recovered in the mock catalogue.  Finally, we repeat this procedure
using the real 2PIGG catalogue, and show the results in filled triangles (groups
with at least $10$ galaxy members, using only $10$ members to measure $b/a$).
It is clear from the figure, that we have been able to reconcile the results
from numerical simulations with those from observational data, as the 2PIGG
groups show slightly more elongated shapes for massive systems 
($b/a\simeq0.53\pm0.023$ for $M =4\ 10^{12}$h$^{-1}M_{\odot}$ and 
$b/a\simeq0.49\pm0.06$ for $M =2\ 10^{14}$h$^{-1}M_{\odot}$, implying 
a $1-\sigma$ detection of a trend).

\begin{figure}
{\epsfxsize=8.truecm 
\epsfbox[40 170 575 705]{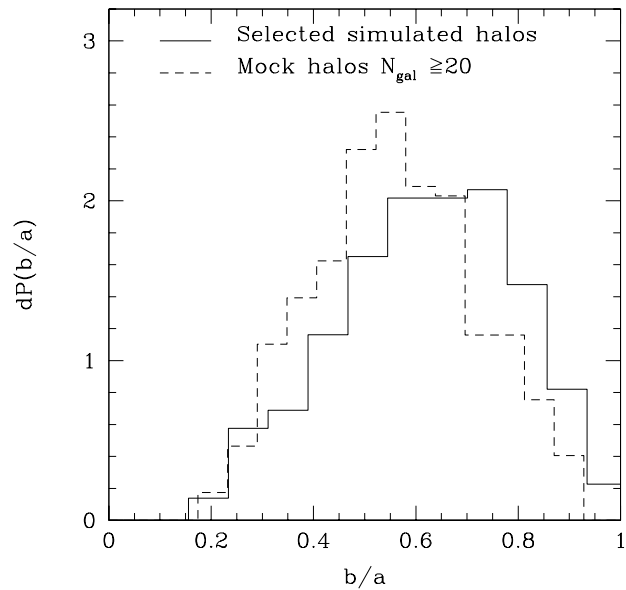}}
  \label{fig:proymock}
  \caption{$b/a$ distributions of projected 3-dimensional haloes (solid line) 
           and of mock groups (doted line). The distributions of
           3-d halo mass and number of member distributions have been matched to those of the 
           mock group sample.}
\end{figure}

\begin{figure}
{\epsfxsize=8.truecm 
\epsfbox[40 170 575 705]{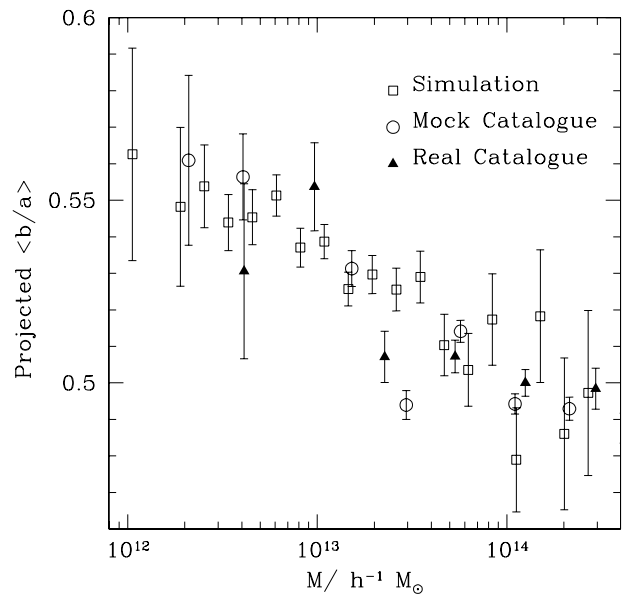}}
  \label{fig:massdep}
  \caption{Dependence of average $b/a$ ratios as a function of mass for
           haloes from the numerical simulation (circles), groups from
           the 2PIGG mock and real catalogues (squares and triangles,
           respectively).  In all cases, only $10$ randomly selected group
           members were used when calculating the projected group shape.}
\end{figure}

\subsection{Dependence of group shape on group member properties}

We now search for possible differences in the typical shapes of groups
when considering different properties of their galaxy members.
We divide our samples of groups according to the average member galaxy colour 
and galaxy spectral type, since these can be considered
indicators of different stellar formation rate histories and
merger or interaction activity within the groups.  
In all comparisons,
we ensure the same number of members is considered when measuring group shapes.

\begin{figure}
{\epsfxsize=8.truecm 
\epsfbox[40 170 575 705]{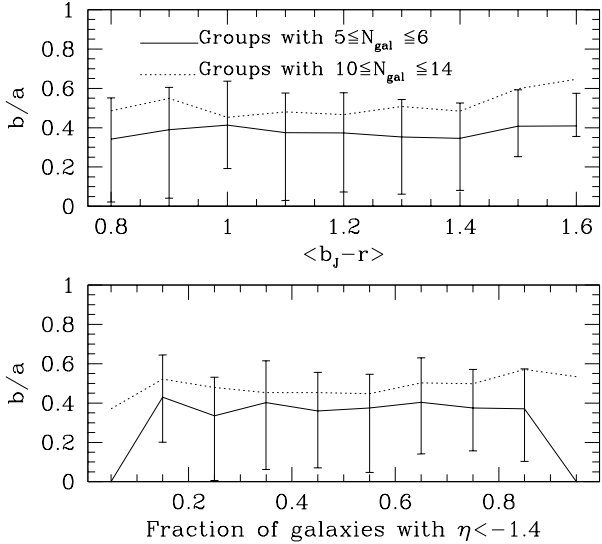}}
  \label{fig:color}
  \caption{Dependence of average $b/a$ ratios as a function of 
           average $b_J-r$ colour (top panel) and fraction of group members with
           $\eta<-1.4$ (bottom panel) for 2PIGG groups.  Solid lines 
           (dotted lines) show the results for groups with 
           $5\le N_{gal}\le 6$ ($10\le N_{gal}\le 14$).  Error-bars 
           correspond to the $10$ and $90$ percentiles.
           }
\end{figure}

In general, we find that it is quite difficult to assess whether a 
correlation between shape and colour is detected in either the SDSS-DR3GC or
2PIGG catalogues.  In particular, we present the results for the 2PIGG groups 
in the top panel of figure 10, 
where we show the median $b/a$ as a function of $b_J-r$ colour, for two
different ranges of number of group members $N_{gal}$ (note the low
number statistics effect inducing overall lower median $b/a$ values for the lower
$N_{gal}$ case).  We repeat this analysis using the spectral type index $\eta$ 
(an indicator of the star formation activity in galaxies) in 2dFGRS galaxies,
determining the fraction of member galaxies that satisfy $\eta<-1.4$ in each
group.  As can be seen in the bottom panel of figure 10, we again find no clear 
dependence of the average semi-axis ratios with
the fraction of passively star forming galaxies within the group.
In the case of SDSS-DR3GG groups, we perform this analysis using the
e-class spectral index, also finding no clear trends.  

We repeated the analysis using only member galaxies with $\eta<-1.4$
(non-star forming galaxies) on the one hand, and only member galaxies 
with $\eta>-1.4$ (star forming galaxies) on the other.   Note
that several groups are represented in both samples, since it is fairly common
that there will be both, star forming and non-star forming galaxies in a group.  
Once more, we were unable to detect any systematic differences in the shapes 
of groups when considering only either star forming or non-star forming galaxies.  
Repeating this analysis using the spectral type index in the SDSS-DR3GC 
catalogue produces the same results, namely, that
there is no clear correlation between spectral type and group shape.

These results are somewhat surprising and indicate the need
for larger samples of groups,
which may provide better means to detect small differences between
group shapes computed considering star forming and non-star forming
galaxies separately.  This could be expected, for instance, if galaxy morphological
segregations in groups were to depend only on the distance to the group centre.

\section{Conclusions}

We have performed several analyses of shapes of gravitationally bound systems
in numerical simulations, mock group catalogues, and observational group catalogues 
derived from the 2dFGRS and the SDSS-DR3 galaxy surveys.
Our main results on the three-dimensional shapes of dark matter haloes in
the numerical simulations can be summarised as follows:
\begin{itemize}
\item Halos are well described by triaxial ellipsoids 
that tend to be more prolate as the halo mass
increases.  This can also be detected by means of other statistical parameters;
both the asphericity $c/a$, and the projected semi-axis ratios $b/a$ 
show a clear decrement with halo mass.
\item We demonstrated that low number statistics tend to bias the
measured shapes toward oblate shapes.
Performing a Monte-Carlo analysis, we show that the smaller semi-axes are more 
affected by noise, and consequently, prolate shapes,
characterised by two small semi-axes, are more distorted.
\end{itemize}

Regarding the analysis of mock and real catalogues, our results can be
summarised as follows:
\begin{itemize}
\item The analysis in redshift-space does not change significantly the
distribution of projected shapes of haloes for a given 
distribution of group mass and number of members, 
although there is a weak tendency for systems in the mock
catalogue to be more elongated than those obtained by direct projection
of the simulations.  This is probably due to differences in the identification
algorithms used in simulations and redshift-space samples.
\item The properties of the distribution of shapes in the mock catalogue
closely resemble those in the observations, indicating the 
accuracy of the GALFORM model in combination with the numerical
simulation to reproduce the observational data.
In particular, the tendency of groups with
low number of members to show elongated shapes
is entirely consistent in both real and mock catalogues.
\item We demonstrate that the observed trend where richer groups are
rounder is an artifact of low number statistics.  We provide
a fit to a corrected distribution of group shapes, calibrated
to match what would be obtained for a fixed $N=20$ members.
\item We also analysed subsamples of 2PIGG and SDSS-DR3GC groups
selected by member galaxy colour 
and spectral type. We were unable to detect a significant dependence of group
shapes on fractions of red galaxies, or fractions of passively star
forming galaxies.
\end{itemize}
Finally, by considering a fixed number of members per group, we were able
to detect a statistically reliable trend of 2PIGG group shapes with mass 
(group masses are obtained via a mass-to-light ratio calibrated by Eke et al., 2004).  
In order to do this, we select 2PIGG groups with at least $10$ member galaxies, and
use only $10$ randomly selected members to measure the group projected semi-axis
ratios.  Our findings indicate that more massive groups tend to show more elongated
shapes, in excellent agreement with results from numerical simulations.

\section*{Acknowledgments}

This work was partially supported by the Concejo Nacional de 
Investigaciones Cient\'{\i}ficas y Tecnol\'ogicas (CONICET), 
the Asociaci\'on Argentina de Astronom\'{\i}a, 
the ESO-Chile Joint Committee, and  
the European Union's ALFA-II
programme, through LENAC, the Latin american European Network for
Astrophysics and Cosmology.
NDP was supported by a Proyecto Postdoctoral Fondecyt
no. 3040038.  We acknowledge helpful discussions with Carlton Baugh.
We thank Vincent Eke for providing the group identification software
used for constructing the 2PIGG group sample.  The numerical simulation
used in this work was kindly provided by the Cosmology Group at the Institute
for Computational Cosmology (Durham).

\end{document}